\documentclass[a4paper,11pt]{article}
\usepackage[margin=0.7in]{geometry}
\usepackage[T1]{fontenc}
\usepackage{amsmath}
\usepackage{amsfonts}
\usepackage{mathtools}
\usepackage{graphicx,caption}
\usepackage{epstopdf}
\usepackage{natbib}
\usepackage{color}

\bibpunct{(}{)}{;}{a}{}{,}
\pdfoutput=1
\usepackage{multirow}
\usepackage{booktabs}

\newlength{\verticalcompensationlength}
\newcounter{verticalcompensationrows}

\begin{document}

   \title{Exploiting periodic orbits as dynamical clues for Kepler and K2 systems}

 \author{Kyriaki I. Antoniadou$^{1,2}$ and Anne-Sophie Libert$^2$\vspace{0.25cm}	\\
				\small{$^1$Department of Physics, Aristotle University of Thessaloniki, 54124,
              Thessaloniki, Greece}\\ 	\small{$^2$NaXys, Department of Mathematics, University of Namur, 8 Rempart de la Vierge, 5000 Namur, Belgium} \\  \small{kyant@auth.gr, anne-sophie.libert@unamur.be}}

\date{}
\maketitle

\begin{center}
							The final publication is available at https://doi.org/10.1051/0004-6361/202037779
							\end{center}
  \begin{abstract}
  Many extrasolar systems possessing planets in mean-motion resonance or resonant chain have been discovered to date. The transit method coupled with transit timing variation analysis provides an insight into the physical and orbital parameters of the systems, but suffers from observational limitations. When a (near-)resonant planetary system resides in the dynamical neighbourhood of a stable periodic orbit, its long-term stability, and thus survival, can be guaranteed. We use the intrinsic property of the periodic orbits, namely their linear horizontal and vertical stability, to validate or further constrain the orbital elements of detected two-planet systems.  
   We computed the families of periodic orbits in the general three-body problem for several two-planet Kepler and K2 systems. The dynamical neighbourhood of the systems is unveiled with maps of dynamical stability.
   Additional validations or constraints on the orbital elements of K2-21, K2-24, Kepler-9, and (non-coplanar) Kepler-108 near-resonant systems were achieved. While a mean-motion resonance locking protects the long-term evolution of the systems K2-21 and K2-24, such a resonant evolution is not possible for the Kepler-9 system, whose stability is maintained through an apsidal anti-alignment. For the Kepler-108 system, we find that the stability of its mutually inclined planets could be justified either solely by a mean-motion resonance, or in tandem with an inclination-type resonance. Going forward, dynamical analyses based on periodic orbits could yield better constrained orbital elements of near-resonant extrasolar systems when performed in parallel to the fitting of the observational data.
 
\end{abstract}
   {\bf keywords}celestial mechanics --
planets and satellites: dynamical evolution and stability -- planetary systems -- methods: analytical -- methods: numerical -- chaos

%
\section{Introduction}

The discoveries made with the NASA Kepler space telescope \citep{keplermission} and later on with the K2 mission \citep{k2mission} radically changed our perception of the universe. Among the $\sim4100$ confirmed exoplanets\footnote{See e.g. {\em exoplanet.eu} \citep{enc} or NASA Exoplanet Archive \citep{nasaconf}.}, the small, rocky, Earth-sized planets have in particular attracted the community's attention in terms of habitability \citep{kop14}. Planetary systems orbiting close to their host star necessitate further studies of the planetary migration and the formation mechanisms (e.g. \citet{war97}, \citet{pap03}, \citet{thommes03}, \citet{hanmu13}, \citet{durkley15}, and \citet{terpap17}). It has been revealed that many transit-detected exoplanetary systems, even the tightly packed ones, possess multiple super-Earth planets locked in mean-motion resonance (MMR) (e.g. \citet{lauchfi02} and \citet{anglada10}) or even in resonant chain (e.g. K2-32 \citep{k232}, K2-138 \citep{k2138}, Kepler-82 \citep{kepler82}, and TRAPPIST-1 \citep{gillon17}). In fact, only a small fraction of the discovered systems are inside the resonant configuration, and an excess of planet pairs is observed just outside the mean-motion resonances (especially the 3/2 and 2/1 MMRs, e.g. \citet{fab14}). 

Due to observational limitations of the transit method, several orbital parameters of the planetary systems cannot be directly inferred from the observations. For the multi-planet systems with resonant pairs of planets, the transit timing variation (TTV) method can additionally provide valuable information on the physical and orbital parameters of the planets (e.g. \citep{egTTV}). Nevertheless, such data are provided with large deviations in some cases, which could have an impact on the long-term (and possibly chaotic) evolution of the planetary system.

As further studies on the long-term evolution and stability of planetary systems require precise knowledge of their orbital elements, dynamical analyses can provide possible dynamical neighbourhoods that match the already-known information about the systems and also favour their survival. The study of the dynamics of a pair of resonant planets revolving around a star is crucial with regard to the long-term stability of such a system. In this respect, the use of the general three-body problem (GTBP) has long been established as an efficient model that can simulate the evolution of planetary systems. Families of periodic orbits form the paths that drive the migration process of the planets \citep[see e.g.][]{bmfm06,hadjvoy10,vat14}, but also shape the stable and unstable domains in phase space, and hence reveal the regions where the exosystems could be hosted. Periodic orbits have proven useful for planetary systems detected with the radial velocity (RV) method. In particular, \citet{a16} unravelled the phase space of the systems HD 82943, HD 73526, HD 128311, HD 60532, HD 45364, and HD 108874 with the use of the resonant periodic orbits close to them. Following the same method, \citet{av16} studied the highly eccentric exoplanets of the systems HD 82943, HD 3651, HD 7449, HD 89744, and HD 102272.

In this work, we focus on several transit-detected extrasolar systems whose pair of planets is evolving close to an MMR, and we compute the families of periodic orbits in the GTBP for each of these systems. The horizontal and vertical stability of the periodic orbits acts as a guide in our quest of regular domains. We aim to provide validations or tighter constraints to fitted orbital elements, and suggestions of possible ranges for unfitted data. In Sect. \ref{model}, we give an insight into the concept of periodic orbits on which our dynamical analysis is based. Our methodology is applied, in Sect. \ref{app}, to the planetary systems K2-21, K2-24, Kepler-9, and Kepler-108. Finally, we summarise our results in Sect.~\ref{con}.

\section{Key points of the dynamical analysis}\label{model}

In our model setup, the  planets and the star are considered as point masses. We adopt a suitable rotating frame of reference in the GTBP (see e.g. \citet[][for the planar case]{hadj06} and \citet[][for the spatial case]{mich79}), and define $\textbf{X}(t)$ as a set of positions and velocities of the two planets. An orbit is periodic if $\textbf{X}(0) = \textbf{X}(T)$, where $T$ is the orbit's period and $t = kT$, with $k \ge 1$ being an integer. Under certain periodicity conditions \citep{avk11}, which determine whether the periodic orbit is symmetric or asymmetric, the system remains invariant. By applying differential corrections and by following specific schemes of mono-parametric continuations, we form the families of periodic orbits. These families shape the stable and unstable domains in phase space, and planets belonging to stable regions can survive long time spans.

Elaborating on the bifurcation and continuation of periodic orbits is out of the scope of this study. However, the reader may refer to \citet{hadj06} and \citet{av12} for the analyses of the different schemes of continuation that are being applied in the GTBP. The horizontal stability of the periodic orbits is based on the eigenvalues of the monodromy matrix of the variational equations of the system \citep[see e.g.][for further details]{hadjbook06}. Along the planar families, we can also define the vertical stability, which creates the vertical critical orbits (v.c.o.) when it changes. These in turn generate spatial families \citep{hen}.

The periodic orbits can either be circular and belong to the circular family, along which the resonance varies  \citep[see][for its presentation]{spis}, or elliptic. In this work, we are interested in the elliptic ones, which are resonant. When viewed in the rotating frame of reference, the periodic orbits correspond to the exact location of the MMR. The MMR of the planets would be $\frac{n_2}{n_1}=\left(\frac{a_1}{a_2}\right)^{-3/2}\approx\frac{p+q}{p}$, where $p, q \in \mathbb{Z}^*$ , and $q$ is the order of the resonance. In this work, subscript 1 (2) always refers to the inner (outer) planet and $a_2$ is normalised to unity. Also, during the computation of the periodic orbits, the total mass of the system and the gravitational constant were normalised to unity.

Moreover, the periodic orbits correspond to the stationary solutions of an averaged Hamiltonian, which are called apsidal corotation resonances (ACRs) \citep{mbf06} and depend on the resonant angles which are, herein, defined as 
\begin{equation}\begin{array}{l}
\theta_1=p\lambda_1-(p+q)\lambda_2+q\varpi_1, \\
\theta_2=p\lambda_1-(p+q)\lambda_2+q\varpi_2,\\
\theta_3=p\lambda_1-(p+q)\lambda_2+\frac{q}{2}(\varpi_1+\varpi_2), \\
\end{array}
\end{equation}
where $\lambda_i=M_i+\varpi_i$ is the mean longitude, $M_i$ the mean anomaly, and $\varpi_i$ the longitude or pericentre ($i=1,2$). If the periodic orbit is symmetric, the resonant angles are equal to 0 or $\pi$, and we can have either aligned ($\Delta\varpi=0$), or anti-aligned ($\Delta\varpi=\pi$) orbits by locating the planets at pericentre or apocentre at $t=0$. The pairs of the resonant angles help us distinguish the four different symmetric configurations that exist. In particular, when $q$ is odd we can use the pair $(\theta_1,\theta_2)$. The arguments of the pairs are $(0,0)$ with $\Delta\varpi=0$, $(0,\pi)$ with $\Delta\varpi=\pi$, $(\pi,0)$ with $\Delta\varpi=\pi$, and $(\pi,\pi)$ with $\Delta\varpi=0$. However, when $q$ is even, we cannot use the same pair of resonant angles, since $\theta_1-\theta_2=q \Delta\varpi$. More specifically, when $q=2$ we use the pair $(\theta_3,\theta_1)$ in order to distinguish the four configurations. When these resonant angles along with the apsidal difference $\Delta\varpi$ rotate, the planets are not in MMR. The libration of these angles can showcase stable domains in phase space, which in each case are linked with a trapping in an MMR, or a secondary resonance, or an apsidal resonance: all of which can guarantee the long-term stability of the system. 

In Hamiltonian systems, stable periodic orbits are surrounded by invariant tori in phase space where the motion is regular \citep{cont}. On the other hand, the vicinity to unstable periodic orbits can cause instability events, such as collisions or escapes. Therefore, the link of real planets with stable domains in their dynamical neighbourhood is fundamental. In order to visualize the phase space, we compute maps of dynamical stability (DS maps) by using the detrended fast Lyapunov indicator (DFLI, \cite{voyatzis08}), as our chaotic indicator, which can be defined as
\begin{equation}                
DFLI(t)=log \left ( \frac{1}{t}\max\{|\mathbf{\xi_1}(t)|,|\mathbf{\xi_2}(t)|\} \right ),
\end{equation}
where $\xi_i$ are the initially orthogonal deviation vectors computed after numerical integration of the variational equations. The DFLI remains almost constant over time for a regular orbit and takes values less than 10. However, the DFLI increases exponentially, taking very large values when chaoticity is traced. In general, a value larger than 15 corresponds to a chaotic orbit. In this study, the maximum integration time for the DFLI is $t_{max}=250\;\rm{kyr}$, which has proven to be an adequate time for the DFLI to reliably distinguish order from chaos. Additionally, the chosen threshold is 30, where we stop the numerical integration and classify the orbit as chaotic. Hence, in the common colour-coding used in each of the DS maps herein, the dark colours correspond to domains where the motion is regular, while the pale ones correspond to regions where the motion is irregular.

\section{Application to Kepler and K2 systems}\label{app}
We dynamically analysed the two-planet systems K2-21, K2-24, Kepler-9, and Kepler-108 with the use of the periodic orbits, in order to validate or further constrain their observational orbital elements, shown in Table \ref{tab1}. The selection of the systems was based on the fact that the planets are (sometimes very) close to an MMR, and their eccentricities are determined as non-zero values, and in particular, they are close to 0.1.

\begin{table*}
\centering
\caption{Published data that were utilised for the study of the two-planet systems.}
\begin{tabular}[b]{lcccc}
\toprule
Parameter&K2-21$^1$&K2-24$^2$&Kepler-9$^3$&Kepler-108$^4$\\
\cmidrule{1-5}
$m_S\left[M_\odot\right]$&$0.676^{+0.06}_{-0.06}$&$1.07^{+0.06}_{-0.06}$&$1.022^{+0.029}_{-0.039}$&$0.96^{+0.29}_{-0.16}$\\
\cmidrule{2-5}
$m_b\left[m_J\right]$&0.01493$^a$/0.01105$^b$&$0.0598^{+0.0069}_{-0.0066}$&$0.137^{+0.005}_{-0.006}$&0.41334\\
\cmidrule{2-5}
$m_c\left[m_J\right]$&0.01978$^a$/0.02063$^b$&$0.0485^{+0.0060}_{-0.0057}$&$0.0941^{+0.0035}_{-0.0041}$&0.20279 \\
\cmidrule{2-5}
$T_b\left[days\right]$&$9.325038^{+0.000379}_{-0.000403}$&$20.88977^{+0.00034}_{-0.00035}$&$19.23891^{+0.00006}_{-0.00006}$&$49.18341^{+0.00033}_{-0.00033}$\\
\cmidrule{2-5}
$T_c\left[days\right]$&$15.501920^{+0.000918}_{-0.000928}$&$42.3391^{+0.0012}_{-0.0012}$&$38.9853^{+0.0003}_{-0.0003}$&$190.353^{+0.017}_{-0.010}$\\
\cmidrule{2-5}
$a_b\left[au\right]$&$0.076^{+0.002}_{-0.003}$&$-$&$0.143^{+0.007}_{-0.006}$&$-$\\
\cmidrule{2-5}
$a_c\left[au\right]$&$0.107^{+0.003}_{-0.004}$&$-$&$0.227^{+0.012}_{-0.008}$&$-$\\
\cmidrule{2-5}
$e_b$&$0.100000^{+0.148654}_{-0.084698}$&$0.06^{+0.01}_{-0.01}$&$0.0609^{+0.0010}_{-0.0013}$&$0.08050$\\
\cmidrule{2-5}
$e_c$&$0.210000^{+0.206770}_{-0.173719}$&$<0.07$&$0.06691^{+0.00010}_{-0.00012}$&$0.13515$ \\
\cmidrule{2-5}
$i_b\left[^{\circ}\right]$&$88.98^{+0.50}_{-0.31}$&$-$&$-$&$90.44$\\
\cmidrule{2-5}
$i_c\left[^{\circ}\right]$&$88.85^{+0.40}_{-0.15}$&$-$&$89.18^{+0.005}_{-0.006}$&$90.40$\\
\cmidrule{2-5}
$\varpi_b\left[^{\circ}\right]$&$34.48^{+101.88}_{-133.73}$&$-$&$-$&$-$\\
\cmidrule{2-5}
$\varpi_c\left[^{\circ}\right]$&$59.96^{+75.67}_{-120.57}$&$-$&$-$&$-$\\
\cmidrule{2-5}
$\omega_b\left[^{\circ}\right]$&$-$&$-$&$357.0^{+0.5}_{-0.4}$&$-151.44$\\
\cmidrule{2-5}
$\omega_c\left[^{\circ}\right]$&$-$&$-$&$167.5^{+0.1}_{-0.1}$&$-74.80$\\
\cmidrule{2-5}
$M_b\left[^{\circ}\right]$&$-$&$-$&$2.6^{+0.5}_{-0.6}$&$-$\\
\cmidrule{2-5}
$M_c\left[^{\circ}\right]$&$-$&$-$&$307.4^{+0.1}_{-0.1}$&$-$\\
\cmidrule{2-5}
$T_{0b}\left[BJD\right]$&$-$&$-$&$-$&665.12261+2454900\\
\cmidrule{2-5}
$T_{0c}\left[BJD\right]$&$-$&$-$&$-$&816.67628+2454900\\
\cmidrule{2-5}
$\Omega_b\left[^{\circ}\right]$&$-$&$-$&$180$&$0.0$\\
\cmidrule{2-5}
$\Omega_c\left[^{\circ}\right]$&$-$&$-$&$179.0^{+0.3}_{-0.1}$&$21.23$\\
\bottomrule
\end{tabular}\\
\begin{flushleft}
Observational values for the four selected systems provided by (1) \citet{dress17}, (2) \citet{petig18}, (3) \citet{bor19}, and (4) \citet{millsfab17}. The planetary masses of K2-21 were yielded through the $^a$mass-radius-density relation of \citet{weissmarcy14} and $^b$the power-law equation of \citet{lismass}.
\end{flushleft}
\label{tab1}
\end{table*}

\subsection{K2-21}\label{k221}
The K2-21 system was discovered by \citet{petig15}, who combined transit detection with TTVs. It was then studied by \citet{cross16} and revalidated by \citet{dress17}. The latter redetermined the transit parameters by performing fits to the K2 photometry and provided the latest data used for our dynamical analysis (see Table \ref{tab1}). 

From the planetary radii provided by the observations, the planetary masses were yielded in two ways. In the first case, they were derived through the mass-radius-density relations of \citet{weissmarcy14}. We found that $m_b=0.01493 {\rm m_J}$ and $m_c=0.01978 {\rm m_J}$. In the second case, the power-law equation of \citet{lismass} was used, which yielded mass values quite similar to the first, namely $m_b=0.01105 {\rm m_J}$ and $m_c=0.02063 {\rm m_J}$.

We note that in the GTBP, it is the mass ratio\footnote{The mass ratio is defined as $\rho=\frac{m_2}{m_1}$, where $m_i$ $(i=1,2)$ are the planetary masses.} of the planets that affects the dynamics, for example, the families of the periodic orbits in our study, and not the individual absolute mass values themselves \citep[see e.g.][]{beau03,voyhadj05}. Hence, we may have a mass ratio equal to $\rho=1.32$ and $\rho=1.86$, according to \citet{weissmarcy14} and \citet{lismass}, respectively. In our study, after having explored both of the mass ratio values, we found the same results with regard to the dynamics of the system, and we herewith present the ones derived via the relations of \citet{weissmarcy14}.

\begin{figure}
\centering
\resizebox{.6\hsize}{!}{\includegraphics{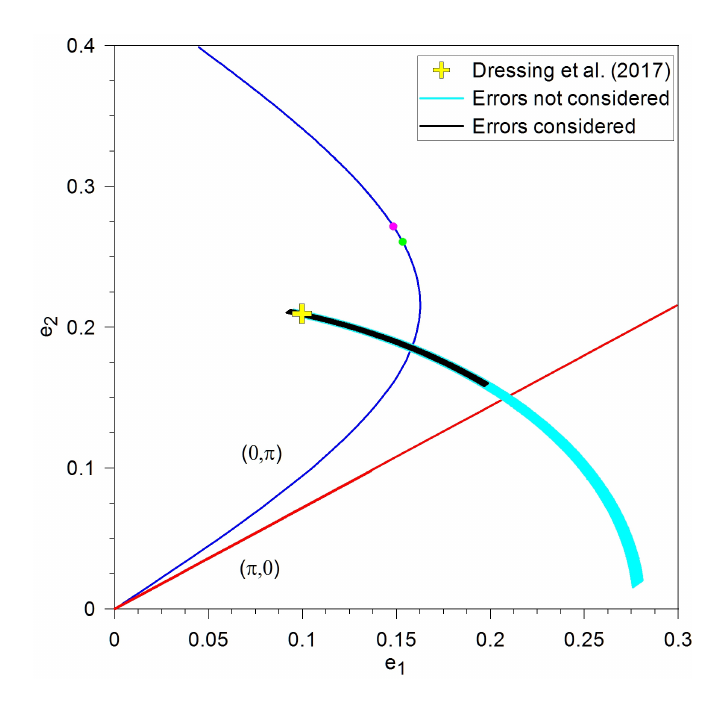}}
\caption{Planar families in the 5/3 MMR computed for planetary mass ratio $\rho=1.32$ of the K2-21 system. The initial conditions (yellow cross) correspond to the eccentricities provided by \citet{dress17}. The cyan curve corresponds to the evolution of the system, when the errors on the orbital elements were not considered. The black curve is centred at a stable (blue) periodic orbit of the family in the configuration $(0,\pi)$, and corresponds to longitudes of pericentre $\varpi_1=135.48{^\circ}$ and $\varpi_2=-60.61{^\circ}$ (values inside the error bars) and mean anomalies $M_1=M_2=0^{\circ}$.}
\label{k221fams}
\end{figure} 

\begin{figure}
\centering
\resizebox{.6\hsize}{!}{\includegraphics{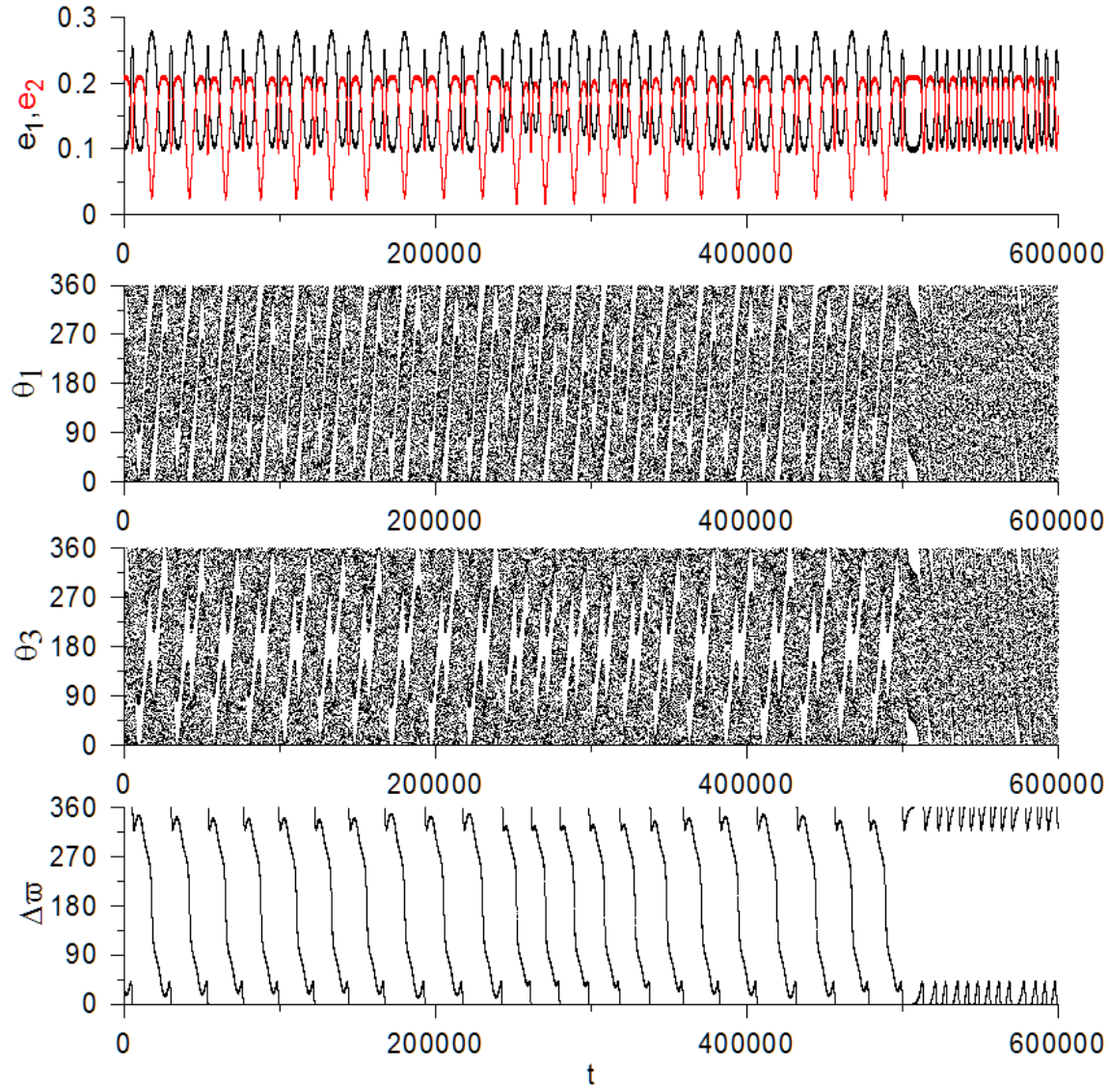}}
\caption{Evolution of K2-21 for observational values given by \citet{dress17}, which corresponds to the cyan curve in Fig.~\ref{k221fams}. The rotation of the resonant angles highlights that the planets are not locked in an MMR, but do rapidly experience a change of dynamical behaviour.}
\label{k221res_dress}
\end{figure}

\begin{figure}
\centering
\resizebox{.6\hsize}{!}{\includegraphics{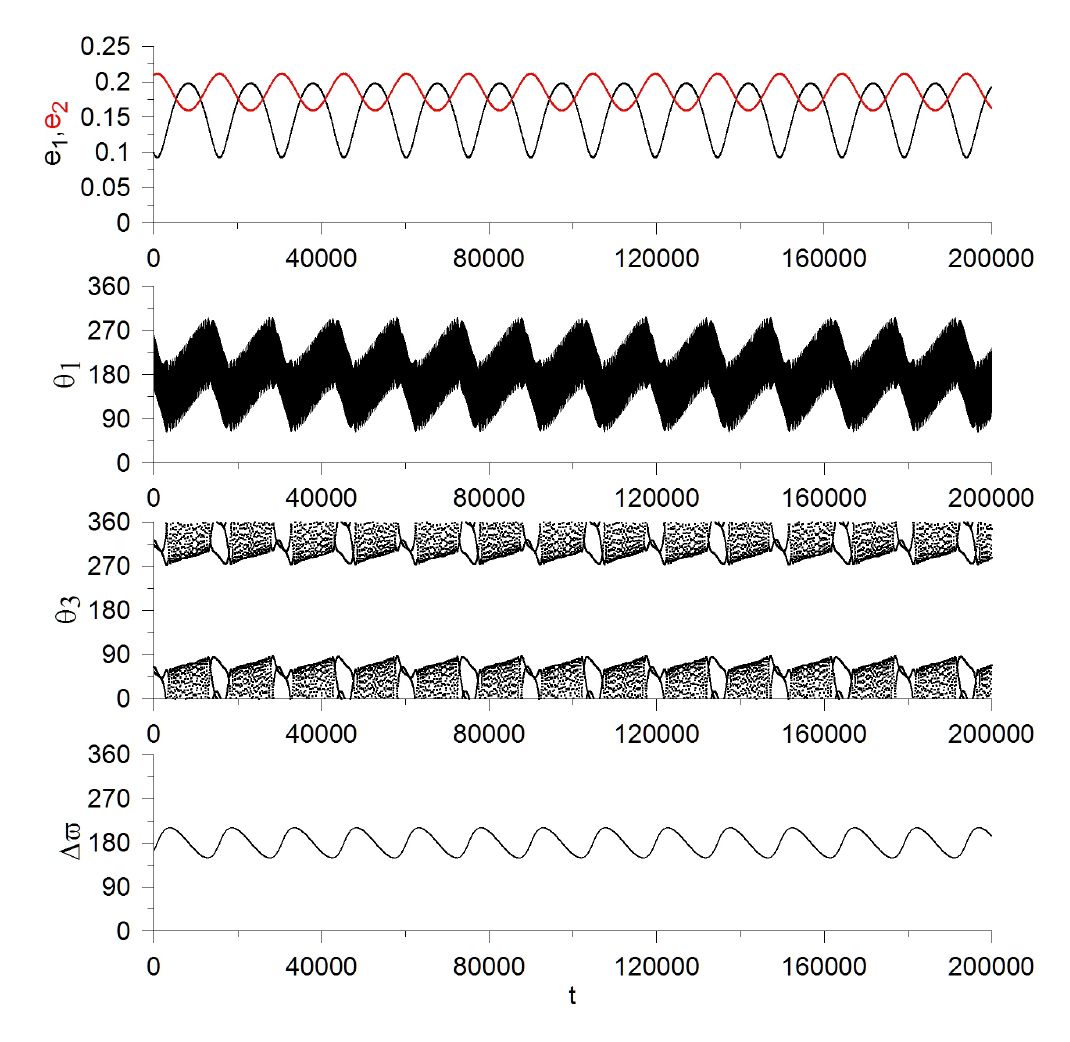}}
\caption{Evolution of K2-21 when errors from \citet{dress17} are taken into account. This resonant evolution (libration of the resonant angles around $(0,\pi)$) corresponds to the black curve in Fig.~\ref{k221fams} and highlights a possible locking in the 5/3 MMR when the orbital elements are selected within the limits of the observational errors.}
\label{k221res}
\end{figure}

According to the period ratio  $\frac{T_c}{T_b}\approx 1.6630$, the system is close to the 5/3 MMR. In Fig. \ref{k221fams}, we present the families of periodic orbits for $\rho=1.32$ in this MMR, which belong to two different configurations: $(\theta_3,\theta_1)=(0,\pi)$ with $\Delta\varpi=\pi$ and $(\theta_3,\theta_1)=(\pi,0)$ with $\Delta\varpi=\pi$. As the mean anomalies are not provided for K2-21, we cannot distinguish the configuration to which the planetary pair belongs, and therefore we cannot but explore all the possibilities in the following.

The evolution of the system, whose initial conditions are indicated by the yellow cross in Fig.~\ref{k221fams}, covers a wide span on the eccentricities plane (cyan curve) and intersects both the family of unstable (red) and the family of stable (blue) periodic orbits. For the nominal orbital elements of \citet{dress17}, the K2-21 system, the evolution of which is shown in Fig.~\ref{k221res_dress}, is not inside, but close to the 5/3 MMR, as indicated by the rotation of the resonant angles. The proximity to the unstable periodic orbits does not guarantee the survival of the planets. Indeed, as shown in Fig.~\ref{k221res_dress}, a change of dynamical behaviour is observed at 0.5 Myr (and 0.2 Myr for $\rho=1.86$) when the apsidal difference starts to oscillate about $0^{\circ}$ (which could protect the phases and stabilise the system if maintained).

Nevertheless, when we selected longitudes of pericentre inside the error bars given by \citet{dress17}, and in particular $\varpi_1=135.48^{\circ}$ and $\varpi_2=-60.61^{\circ}$, the evolution (black curve) centred at a stable (blue) periodic orbit of the configuration $(0,\pi)$. The resonant evolution is shown in Fig.~\ref{k221res} (see the libration of the resonant angles about $(0,\pi)$). It highlights that a possible locking in the 5/3 MMR is possible for K2-21, which would guarantee a long-term stable planetary evolution.

\begin{figure}
\centering
\resizebox{.6\hsize}{!}{\includegraphics{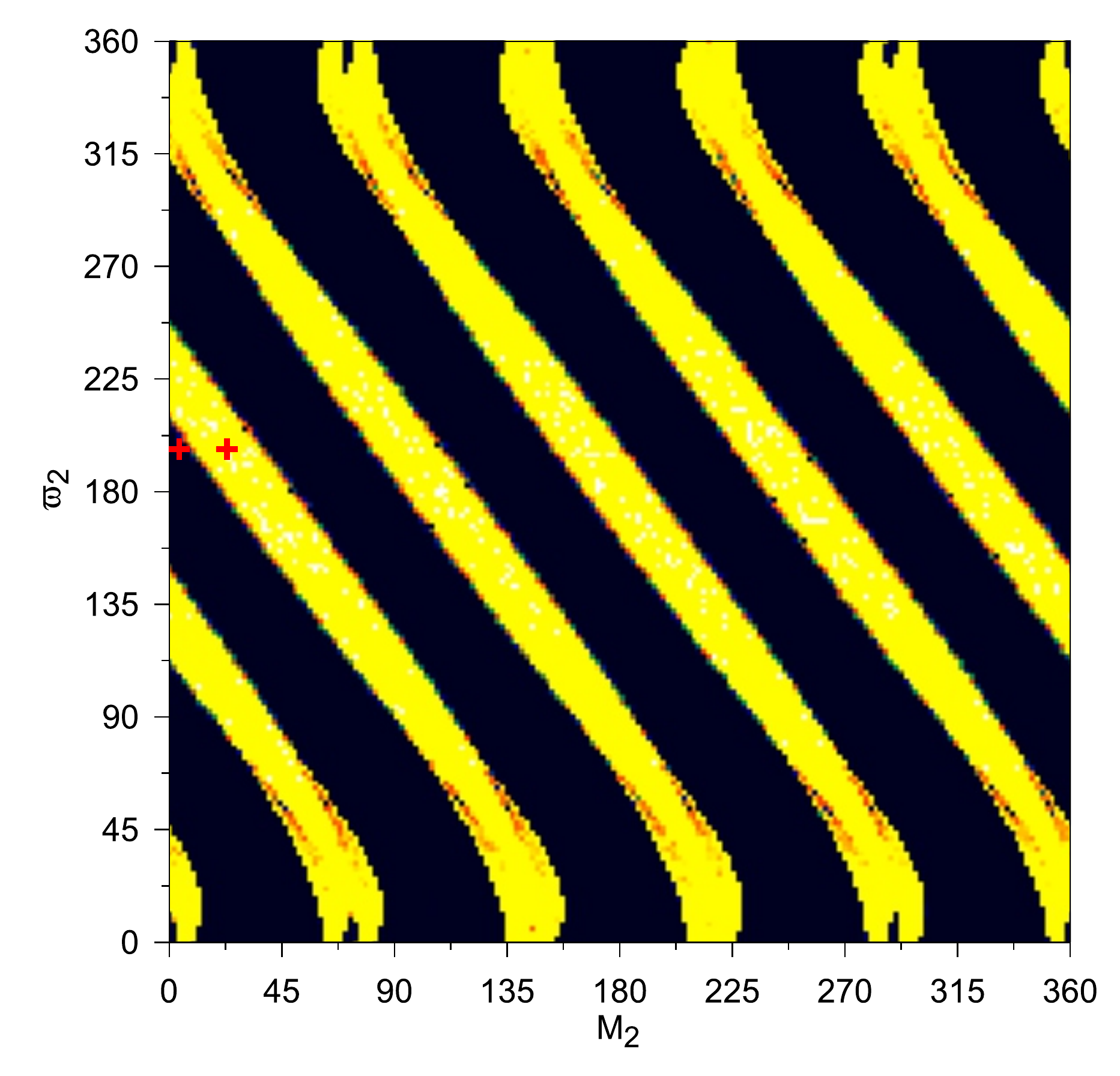}}
\resizebox{.4\hsize}{!}{\includegraphics[width=4.9cm,height=1.1cm]{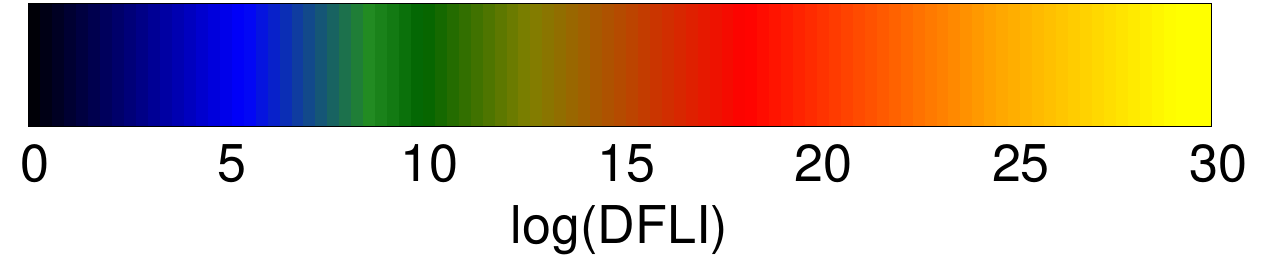}} 
\caption{DS map on $(M_2,\varpi_2)$ plane for K2-21. The long-term stability is guaranteed in the dark-coloured regions. Dark (pale) domains correspond to stable (chaotic) domains in phase space as ascribed by the logarithmic values of the DFLI (coloured bar). The red crosses represent the initial conditions of the evolutions highlighted in Figs. \ref{k221maps} and \ref{k221mapu}.}
\label{k221map}
\end{figure}

\begin{figure}[!ht]
\centering
\resizebox{.6\hsize}{!}{\includegraphics{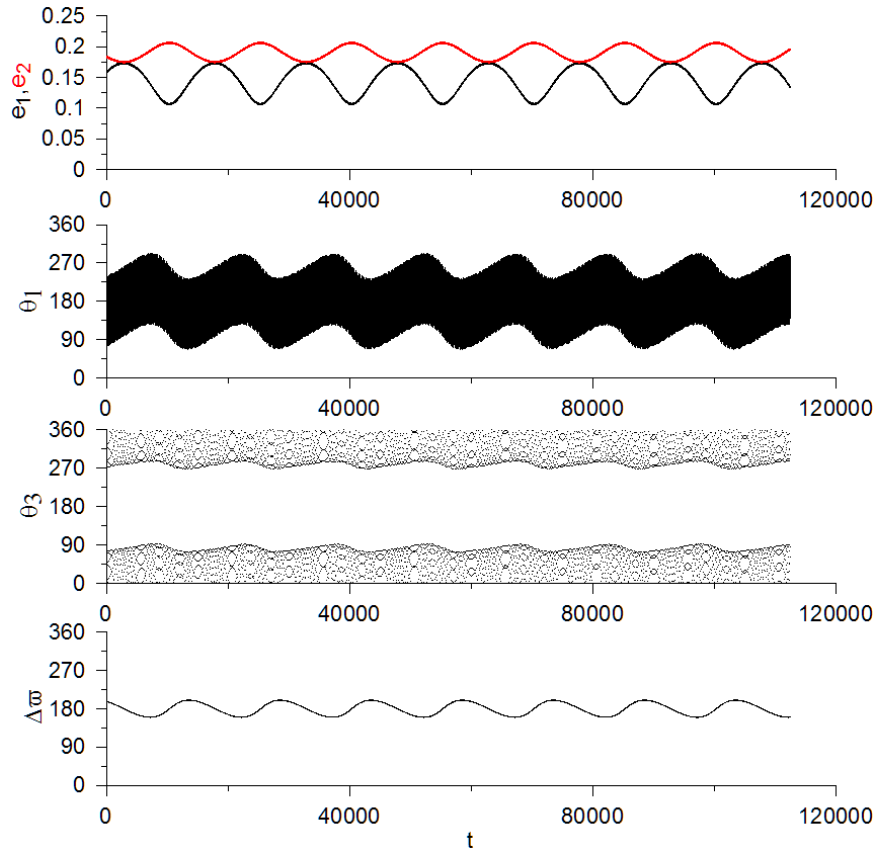}}
\caption{Stable evolution, where resonant angles and apsidal difference librate, from the initial condition (red cross) in the regular (dark) domain of Fig. \ref{k221map}.}
\label{k221maps}
\end{figure}

\begin{figure}
\centering
\resizebox{.6\hsize}{!}{\includegraphics{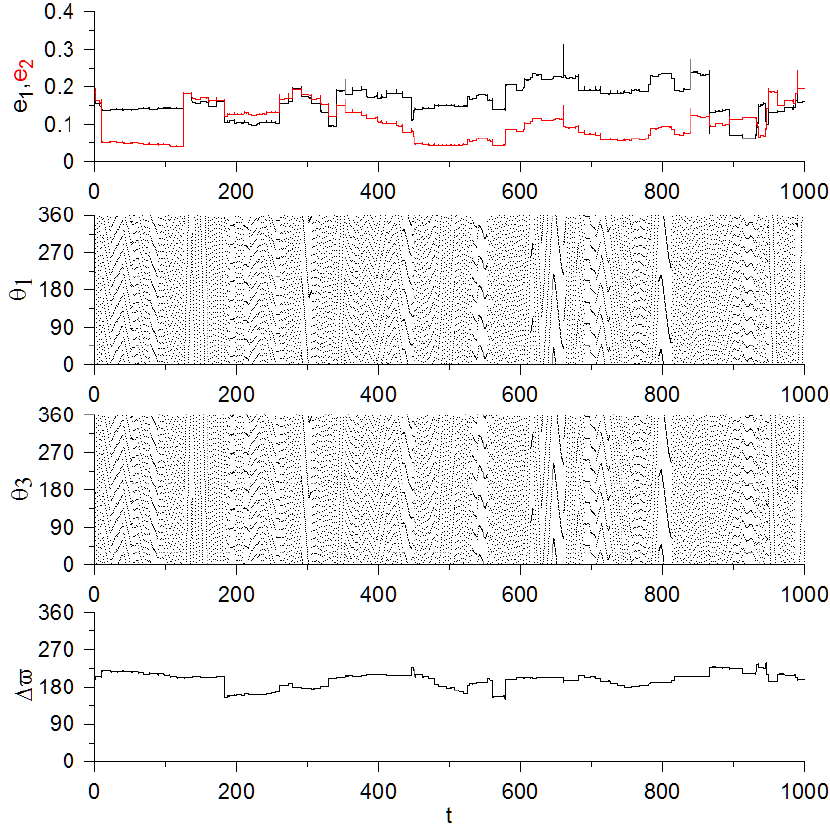}}
\caption{Unstable evolution, where resonant angles rotate, from the initial condition (red cross) in the chaotic (yellow) domain of Fig. \ref{k221map}.}
\label{k221mapu}
\end{figure}

We note that the mean anomalies were fixed to $0^{\circ}$. This choice was not arbitrary, since these values are the ones along the family of stable periodic orbits, where the evolution centred. However, since in nature the values of the mean anomalies are unlikely to be exactly equal to zero, we performed an additional study to explore the influence of these values, as well as the values of the longitudes of the pericentre, on the dynamics of the K2-21 system. We selected the planar periodic orbit appearing on the black evolution in Fig.~\ref{k221fams} with orbital elements $a_1/a_2=0.71148$, $e_1=0.1577$, $e_2=0.184$, $\varpi_1=0^{\circ}$, $\varpi_2=180^{\circ}$ , and $M_1=M_2=0^{\circ}$, and created the DS map on the $(M_2,\varpi_2)$ plane shown in Fig. \ref{k221map}, where the long-term stability is guaranteed in the dark-coloured regions. A similar structure for the 5/3 MMR was observed by \citet{forgacs}.

A great sensitivity of the stability of the system depending on the values of the angles is observed in Fig. \ref{k221map}. To stress the great importance of the angles, we provide a clear example by choosing two different initial conditions (red crosses) from Fig. \ref{k221map}; one located in the domain of regular motion (dark) with $\varpi_2=197^{\circ}$ and $M_2=4^{\circ}$ and another in the irregular one (yellow) with $\varpi_2=199^{\circ}$ and $M_2=22^{\circ}$. In the former case, we see that the resonant angles librate (Fig. \ref{k221maps}) and the stability is guaranteed by the locking in the MMR, whereas in the latter case, the resonant angles rotate and the system is very soon destabilised (Fig. \ref{k221mapu}).

To summarise, we exemplified that the evolution about a stable periodic orbit in addition to a planetary trapping in an MMR could protect the long-term stability of the K2-21 system. Such an evolution was found for orbital parameters inside the error bars of \citet{dress17}. Moreover, within the DS maps, we showed how the extent of the unfitted orbital elements could be concluded, so that long-term stability is favoured. These results are not sensitive to slight changes in mass ratio, and thus they do not depend on the mass-radius relationship considered.

\subsection{K2-24}\label{k224}
The planetary system K2-24 was discovered by \citet{petig16} and has also been studied by \citet{dai16}, \citet{sinu16}, \citet{cross16}, and \citet{mayo}. The latest data for this system are provided by \citet{petig18}, who implemented TTV and RV analyses in order to tightly constrain the eccentricities (see Table \ref{tab1}). 

\begin{figure}
\centering
\resizebox{.6\hsize}{!}{\includegraphics{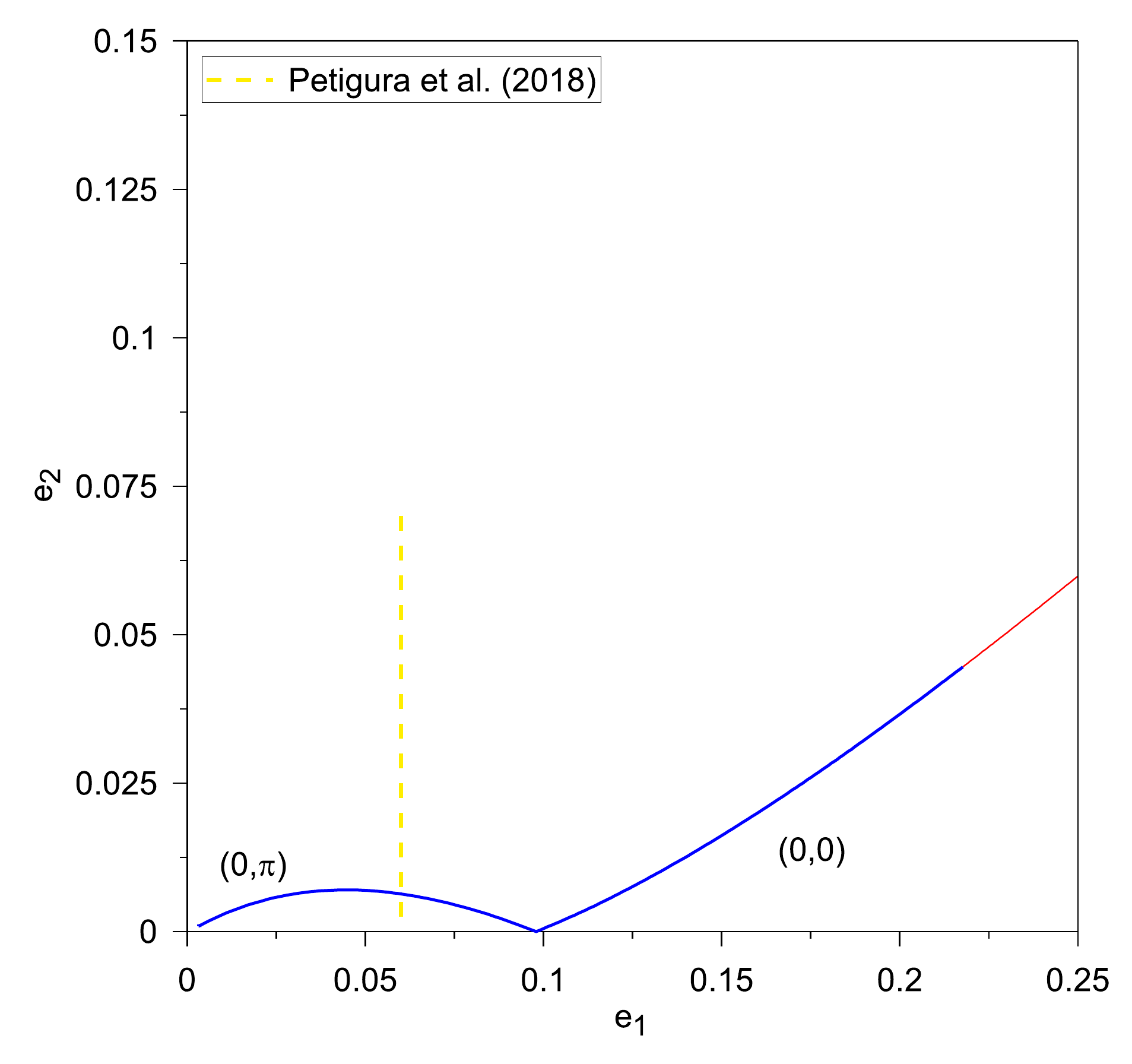}}
\caption{Planar families in 2/1 MMR for planetary mass ratio $\rho=0.8$ of K2-24. Stable (blue) periodic orbits exist in the neighbourhood of the planets, namely for $e_b=0.06$ and $e_c<0.07$ (yellow dashed curve).}
\label{k224fams}
\end{figure}

Given the period ratio $\frac{T_c}{T_b}\approx 2.0268$, the two planets are residing close to the 2/1 MMR, with a mass ratio of 0.8. In Fig. \ref{k224fams}, we present the families of periodic orbits close to the observational eccentricities, namely for $e_b=0.06$ and $e_c<0.07$ (yellow dashed curve). We observe that stable periodic orbits exist for the configurations $(\theta_1,\theta_2)=(0,0)$ with $\Delta\varpi=0$ and $(\theta_1,\theta_2)=(0,\pi)$ with $\Delta\varpi=\pi$. In particular, stable periodic orbits exist when $e_2<0.05$ in the configuration $(0,0)$. Interestingly it almost entirely agrees with the limitation given by the observations $(e_c<0.07)$.

\begin{figure}
\centering
\resizebox{0.6\hsize}{!}{\includegraphics{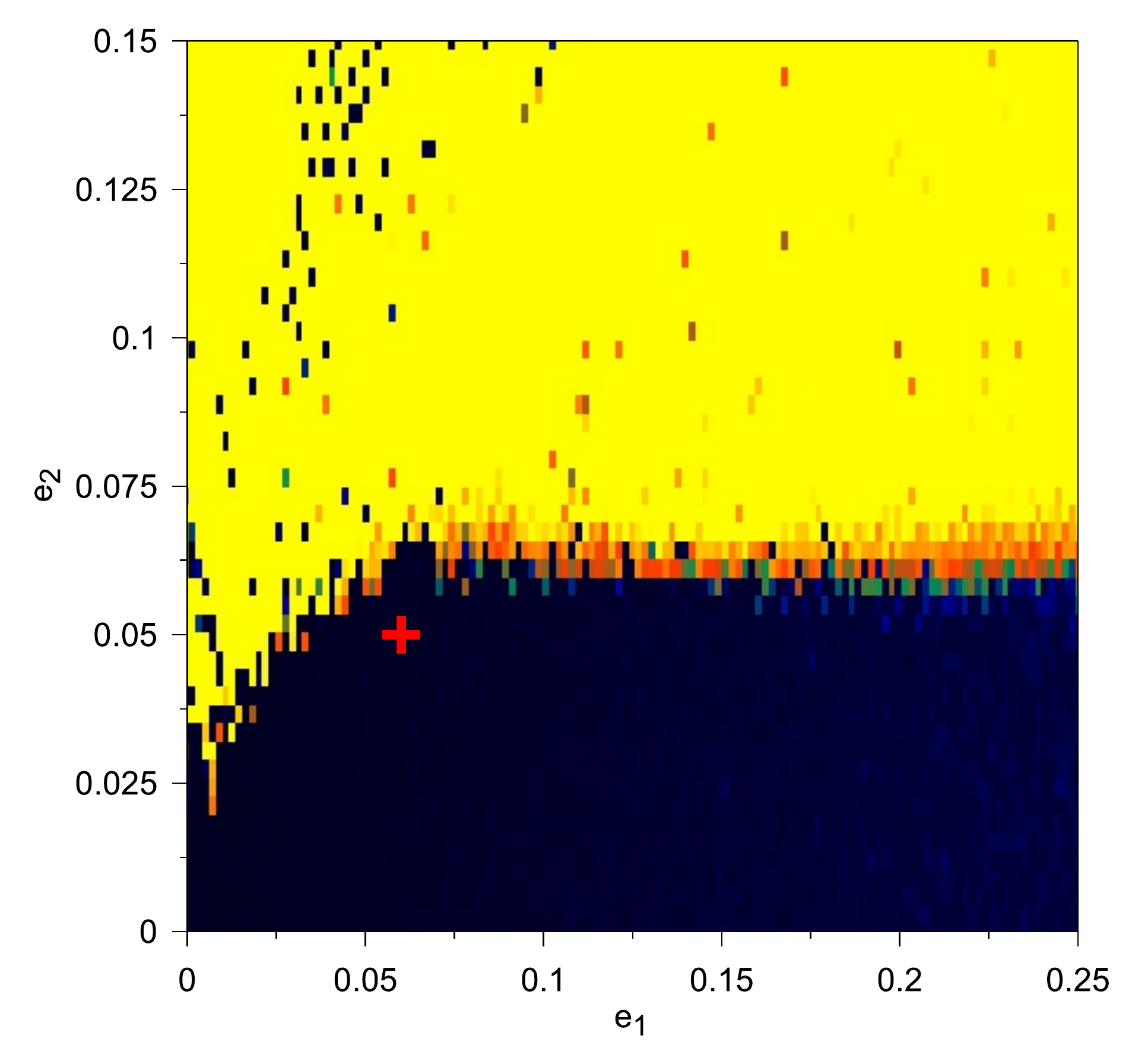}}

\resizebox{.4\hsize}{!}{\includegraphics[width=4.9cm,height=1.1cm]{bar.pdf}} 
\caption{DS map on $(e_1,e_2)$ plane for K2-24. It is shown that the limit condition $e_c<0.07$ of \citet{petig18} is even tighter, since the eccentricity of the outer planet, $e_2$, has to be lower than 0.05.}
\label{k224map}
\end{figure}

By selecting the stable periodic orbit with $e_1=0.06=e_b$ (value of the eccentricity of the inner planet), we computed a DS map on the $(e_1,e_2)$ plane in order to reveal the boundaries that guarantee regular motion in the dynamical neighbourhood of these planets. The rest of the orbital elements of the planar periodic orbit considered are: $a_1/a_2=0.629883$, $e_2=0.0063$, $\varpi_1=0^{\circ}$, $\varpi_2=180^{\circ}$ , and $M_1=M_2=0^{\circ}$. In Fig. \ref{k224map}, we observe that the observational limit for $e_c<0.07$ is somewhat confirmed by the analysis of the DS map, as it has to be constrained to $<0.05$.

\begin{figure}
\centering
\resizebox{.6\hsize}{!}{\includegraphics{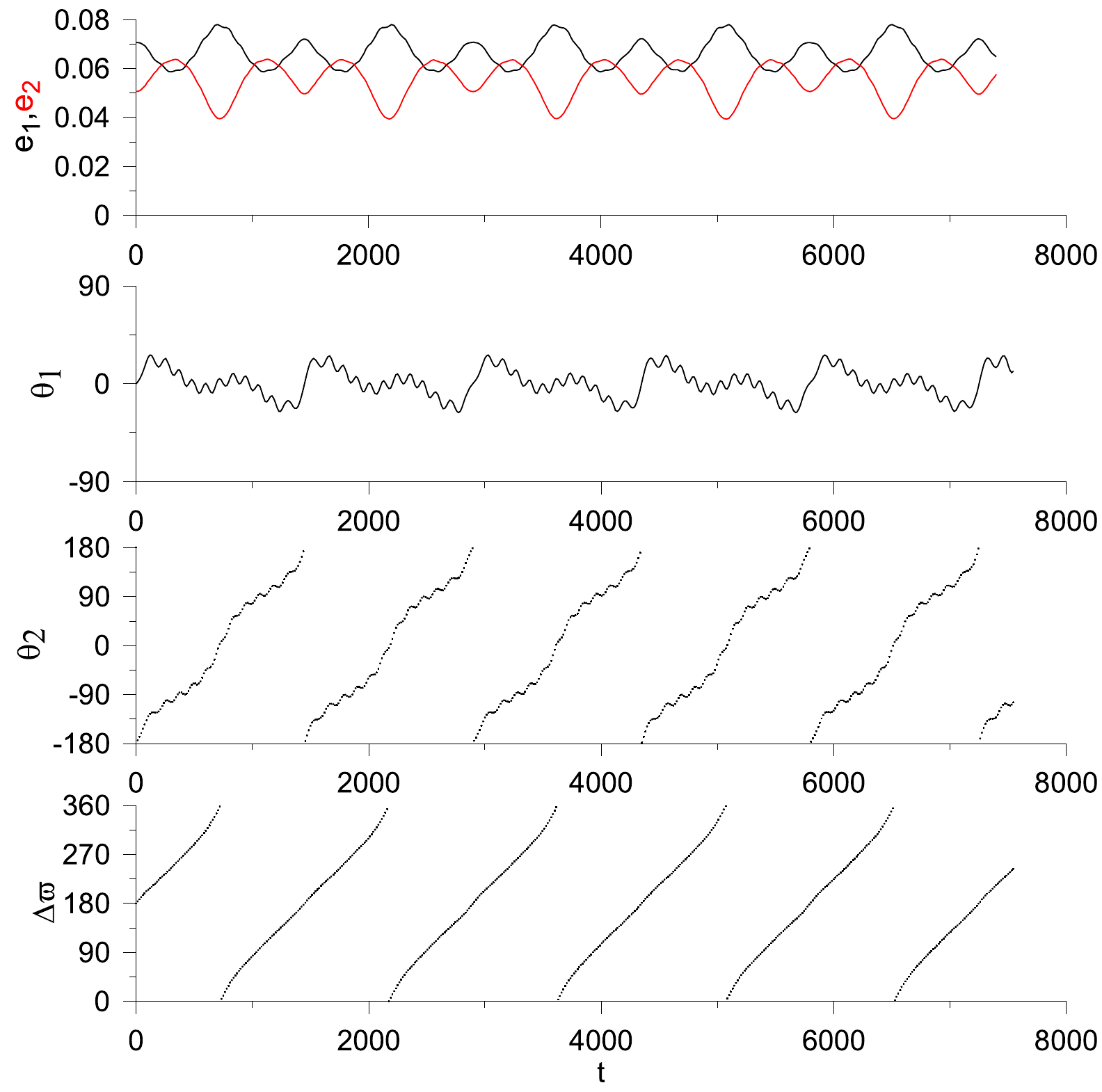}}
\caption{Possible evolution for K2-24 with initial conditions being symbolised by the red cross in Fig.~\ref{k224map}. The 1/1 secondary resonance inside the 2/1 MMR protects the phases of the planets.}
\label{k224evol}
\end{figure}

In order to showcase a stable evolution of the planets within these boundaries, we selected initial conditions from the DS map in Fig. \ref{k224map}. We chose $e_1=0.06$ and $e_2=0.05$, (red cross), the latter value being the upper limit of $e_c$, according to our dynamical analysis. The evolution of the resonant angles of the system is presented in Fig.~\ref{k224evol}. We observe that it is the 1/1 secondary resonance (the average of the frequencies between the rotation of $\theta_2$ and the libration of $\theta_1$) inside the 2/1 MMR that safeguards the planets in this area \citep[see also Fig. 9b in][for more details]{kiaasl}. 

\begin{figure}
\centering
\resizebox{0.6\hsize}{!}{\includegraphics{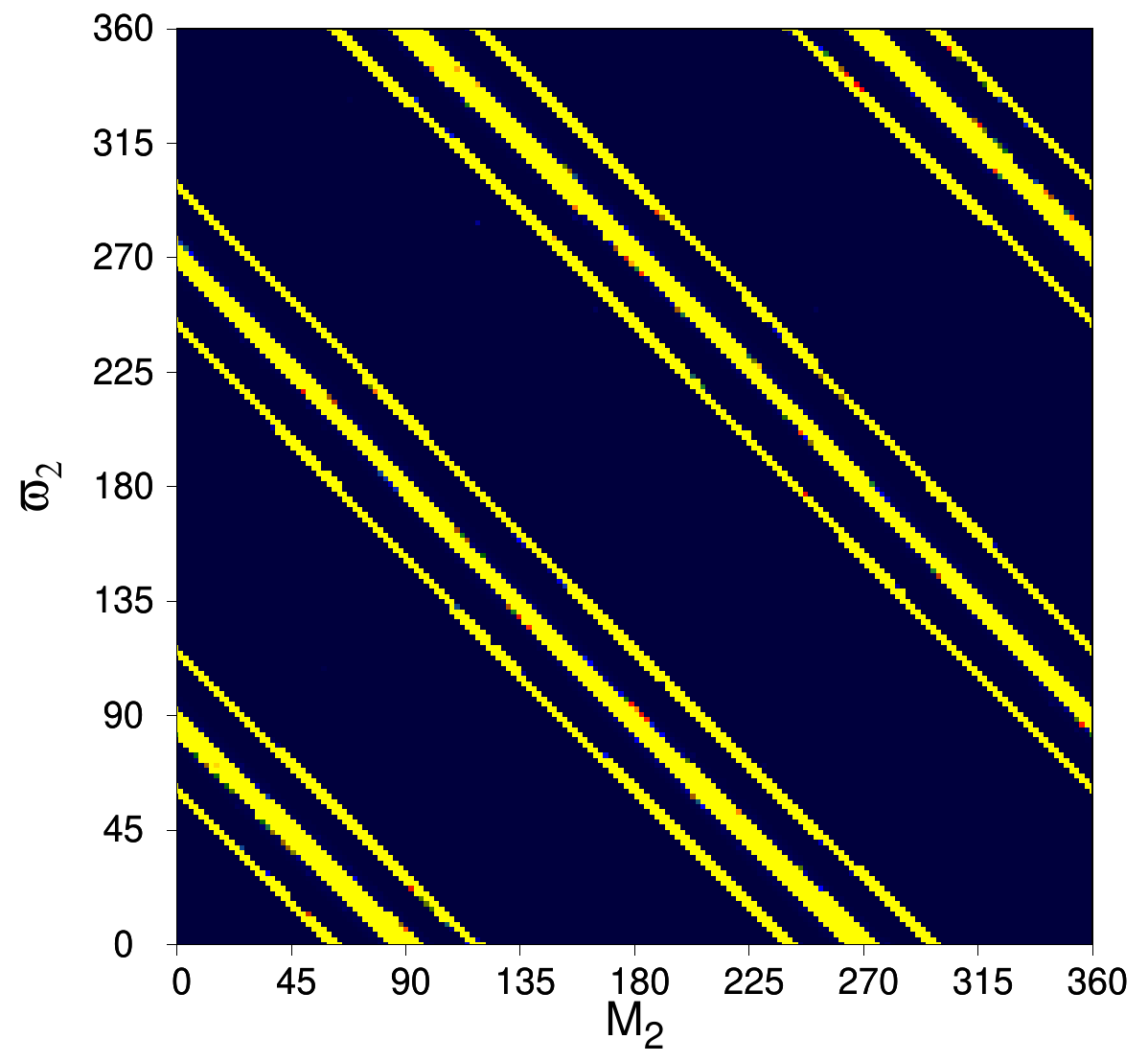}}

\resizebox{.4\hsize}{!}{\includegraphics[width=4.9cm,height=1.1cm]{bar.pdf}} 
\caption{DS-map on $(M_2,\varpi_2)$ plane for K2-24 showing the large extent of the regular domains regarding these angles.}
\label{k224mapMw}
\end{figure}

In Fig. \ref{k224mapMw}, we used the same stable periodic orbit and constructed a DS map on the $(M_2,\varpi_2)$ plane to identify possible values for these angular variables. We highlight that the regular domains that guarantee long-term stability for the K2-24 system are very broad.

Here, we confirmed and further constrained the present-day eccentricities of the K2-24 system introduced by \citet{petig18}. The periodic orbits unravelled the stable region around the system and how the secondary resonance inside the MMR protects the phases of the system.

\subsection{Kepler-9}\label{kepler9}
The Kepler-9 system  was discovered by \citet{hol10} and has been widely studied \citep{bor14,had14,holcz16,mor16,had17,berg18,freu18,Wang182,wang1873,bor19}. \citet{bor19} implemented TTV and RV analyses and provided the data used in our study (see Table \ref{tab1}).

\begin{figure}
\centering
\resizebox{.6\hsize}{!}{\includegraphics{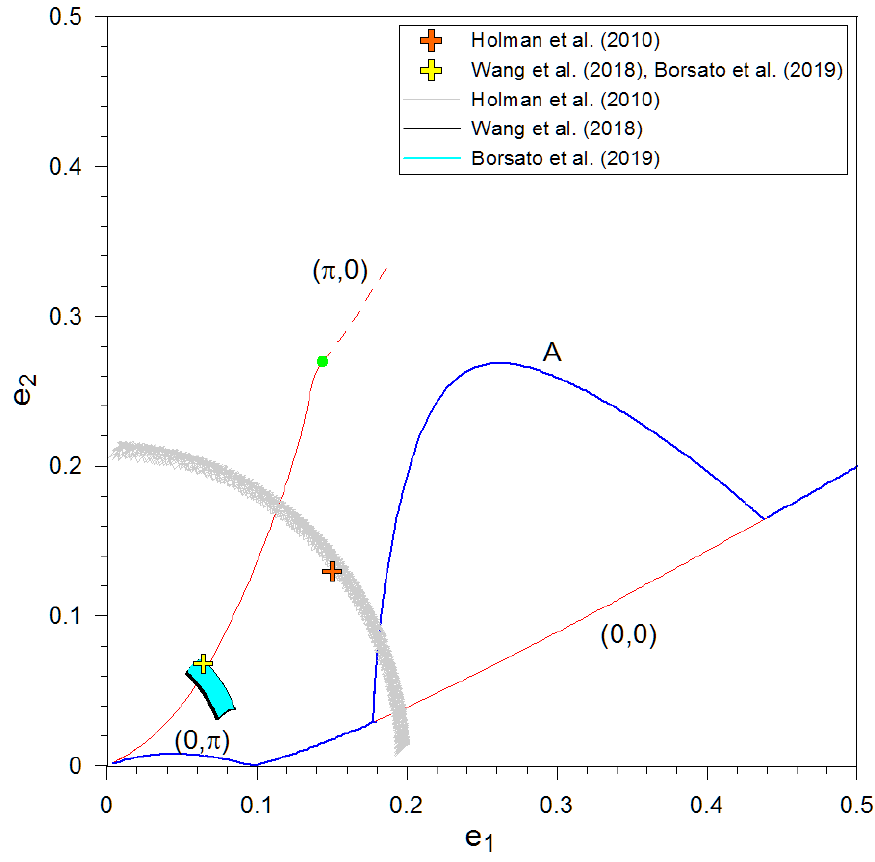}}
\caption{Families of periodic orbits in 2/1 MMR for planetary mass ratio $\rho=0.69$ of  Kepler-9bc, along with the evolutions initiated by the orbital elements given by \citet{hol10}, \citet{wang1873}, and \citet{bor19}, which locate the pair close to the family of unstable (red) periodic orbits.}
\label{kep9fams}
\end{figure}

The b and c pair of Kepler-9 evolves near the 2/1 MMR, since $\frac{T_c}{T_b}\approx 2.0779$. In Fig. \ref{kep9fams}, we present the families of periodic orbits in the 2/1 MMR for $\rho=0.69$. It is made clear that the latest fittings (e.g. \citet{wang1873} and \citet{bor19}) locate the planets close to the unstable (red) periodic orbits (yellow cross in Fig.~\ref{kep9fams}). This explains why Kepler-9 is not inside the 2/1 MMR. Indeed, if the planets were locked in the 2/1 MMR (exactly centred at the family), they would eventually experience instability events.

\begin{figure}
\begin{center}
\includegraphics[width=.6\columnwidth]{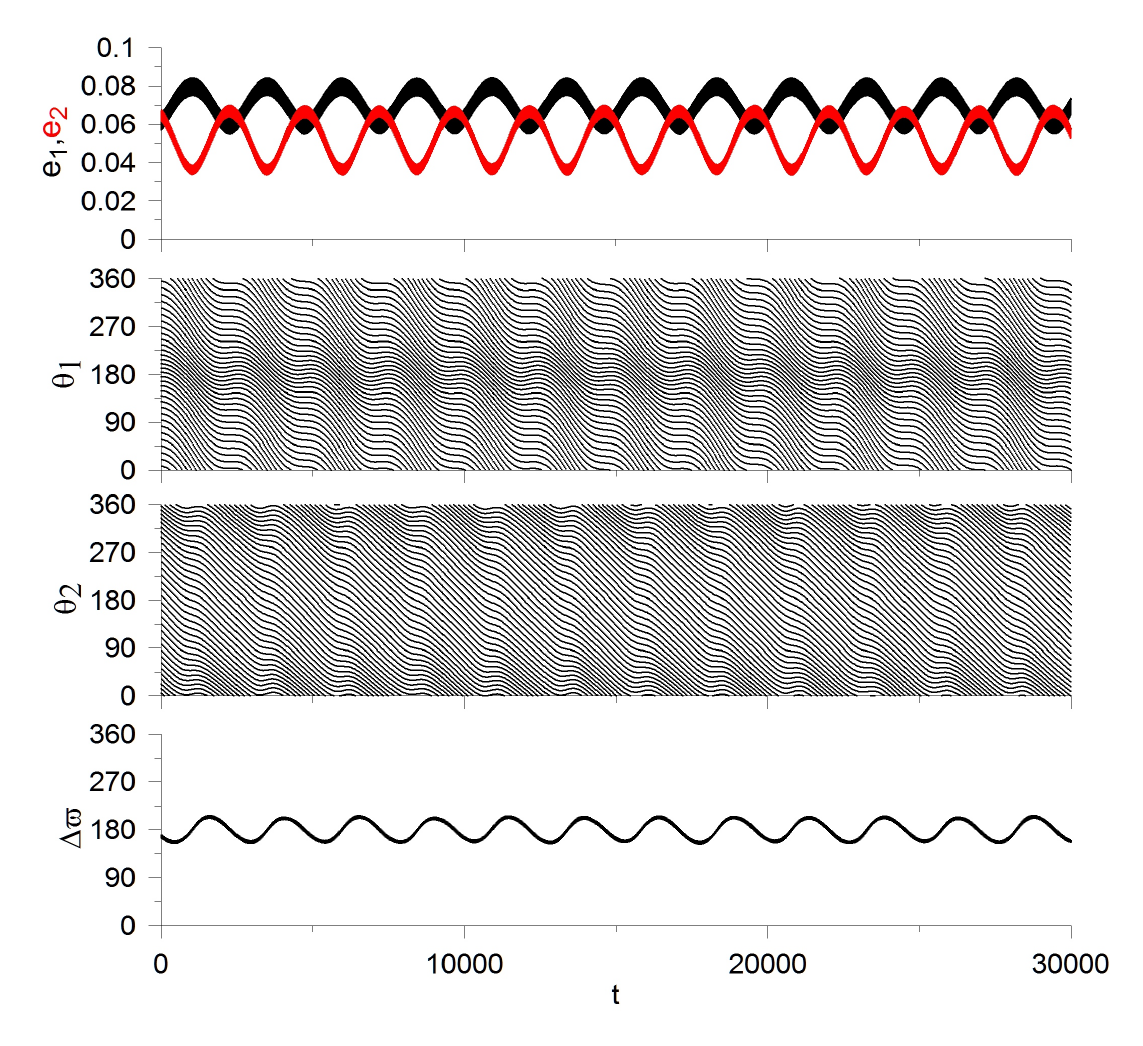}\vspace{-0.4cm}
\caption{Evolution of Kepler-9bc for orbital elements provided by \citet{bor19}.}
\label{kep9res}
\end{center}
\end{figure}

In Fig. \ref{kep9res}, we provide the evolutions of the resonant angles and the apsidal difference based on the latest data provided by \citet{bor19} (cyan evolution in Fig. \ref{kep9fams}). The evolutions of \citet{hol10} and \citet{wang1873} are dynamically the same. The long-term stability is governed by the apsidal difference oscillation about $180^{\circ}$.

Here, we show that the planetary b and c pair of Kepler-9 lies close to but not inside the 2/1 MMR, due to the existence of a family of unstable periodic orbits. The long-term stability is maintained through the oscillation of the apsidal difference.

\subsection{Kepler-108}\label{kepler108}
Kepler-108 was discovered by \citet{rowe14} and since then, has been studied by \citet{eylen15}, \citet{mor16}, \citet{holcz16}, \citet{mustill17}, and \cite{gajdos}. The latest data for our study (see Table \ref{tab1}) were taken by \citet{millsfab17}, who performed photodynamic TTV and transit duration variation (TDV) analyses. The parametrisation they favoured is a highly mutually inclined ($\sim 24^\circ$) giant planet system.

\begin{figure}
\centering
\resizebox{.6\hsize}{!}{\includegraphics{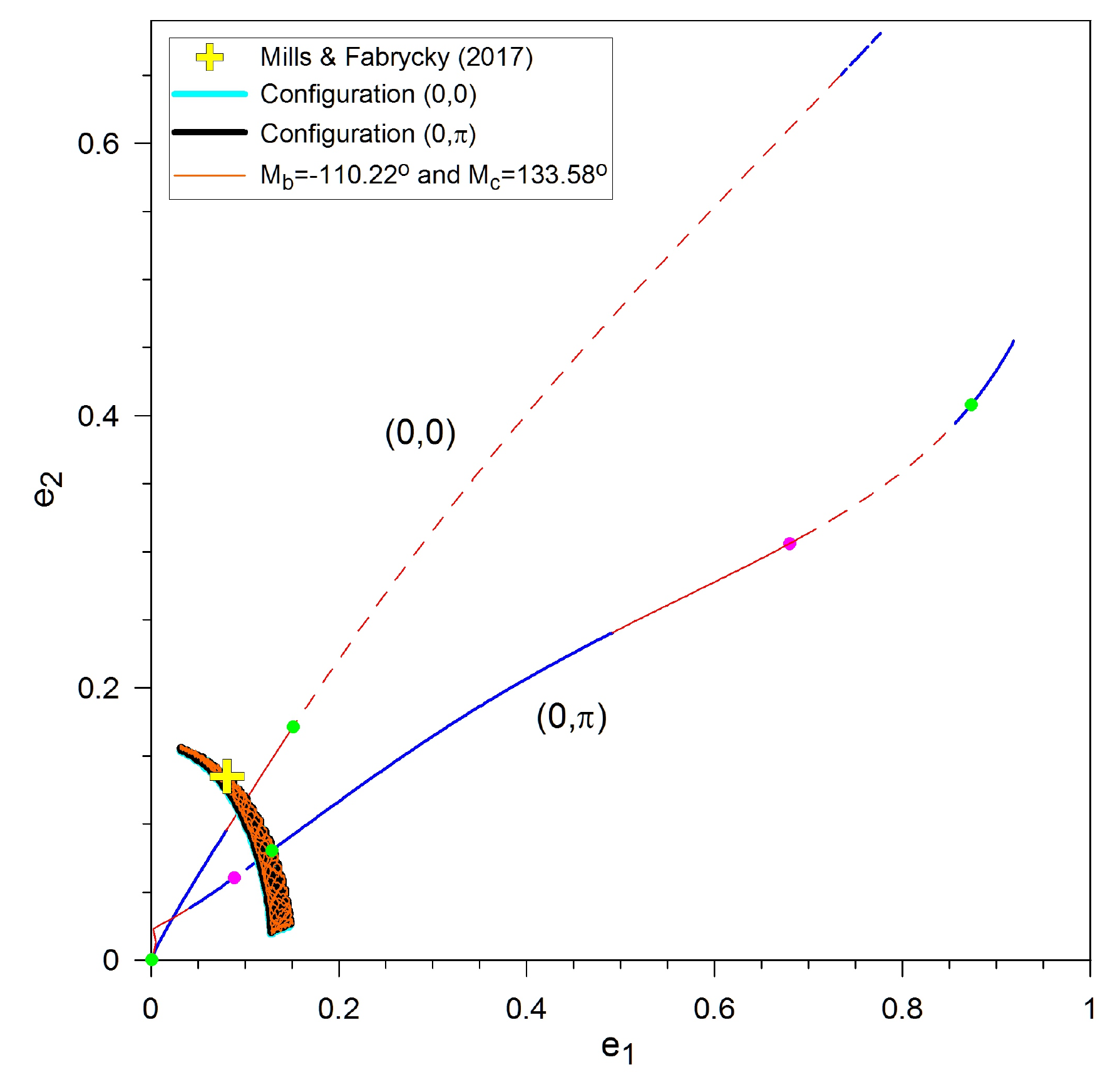}}
\caption{Planar families of periodic orbits in 4/1 MMR for planetary mass ratio $\rho=0.49$ of Kepler-108, which evolves close to this MMR. The black evolution corresponds to the mean anomalies taken by planar configuration ($0,\pi$), while the cyan one is yielded by the angles of the configuration (0,0). The orange evolution corresponds to the calculated observational mean anomalies $M_b=-110.226^{\circ}$ and $M_c=133.58^{\circ}$.}
\label{kep108fams}
\end{figure}

\begin{figure}
\includegraphics[width=.6\columnwidth]{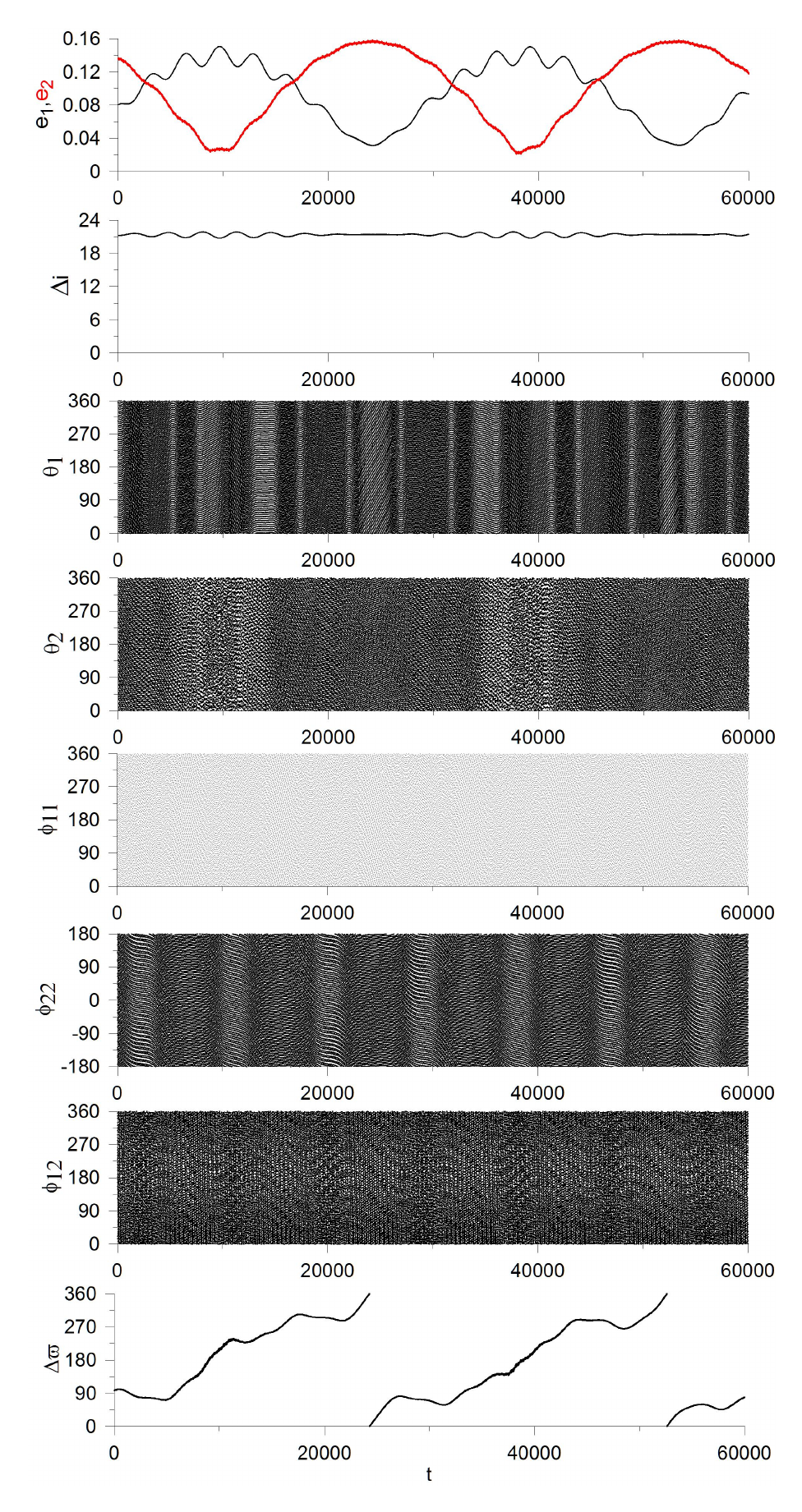}
\caption{Evolution of Kepler-108 for $M_b=-110.226^{\circ}$ and $M_c=133.58^{\circ}$ (orange evolution of Fig.~\ref{kep108fams}). The system is neither captured in the 4/1 MMR, nor in an inclination-type resonance.}
\label{kep108res}
\end{figure}

The period ratio $\frac{T_c}{T_b}\approx 3.8704$ shows that the system is near the 4/1 MMR. As in the previous cases, we computed the planar families of periodic orbits close to the observational values of the eccentricities of the planets. In Fig.~\ref{kep108fams}, we present the ones possessing segments of stable periodic orbits in this vicinity. There are two main families, one belonging to the configuration $(\theta_1,\theta_2)=(0,0)$ with $\Delta\varpi=0,$ and another to the configuration $(\theta_1,\theta_2)=(0,\pi)$ with $\Delta\varpi=\pi$.
\begin{figure}
\centering
\resizebox{.6\hsize}{!}{\includegraphics{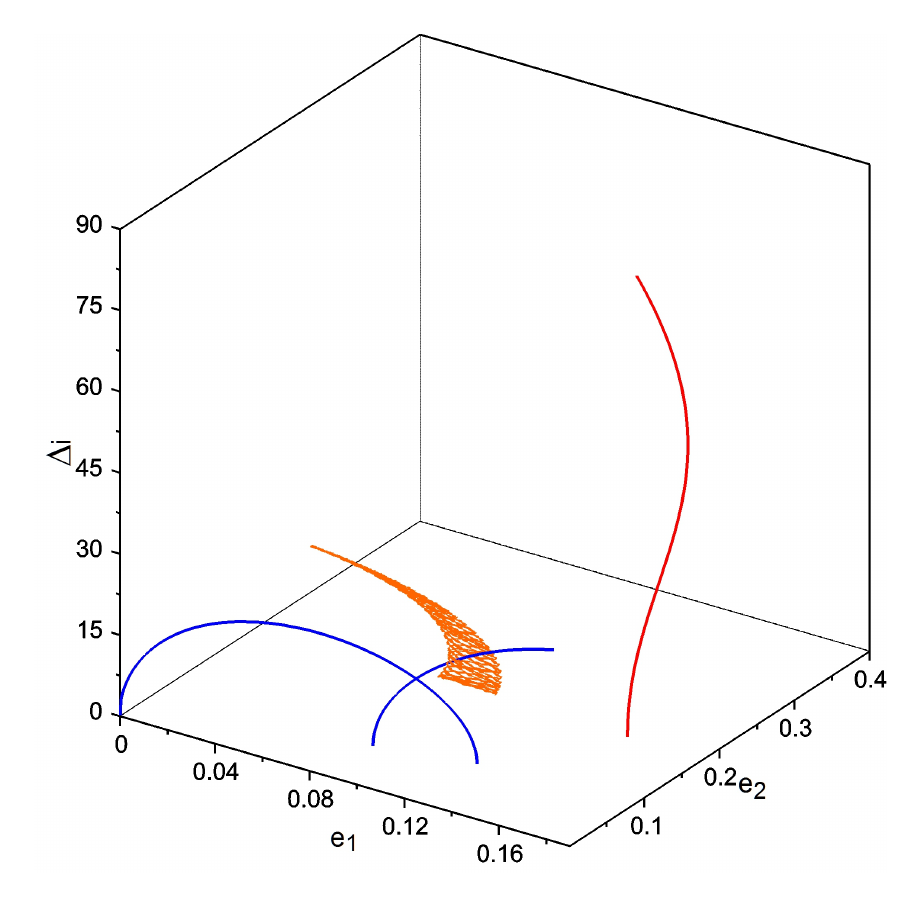}}
\caption{Spatial families of periodic orbits presented in $(e_1,e_2, \Delta i)$ space, which are generated by the v.c.o. (green and magenta dots) shown in Fig. \ref{kep108fams} for Kepler-108. The orange evolution with $M_b=-110.226^{\circ}$ and $M_c=133.58^{\circ}$ is overplotted and crosses a stable (blue) spatial periodic orbit with $\Delta i=24^{\circ}$.}
\label{kep1083dfams}
\end{figure}

\begin{figure}
\begin{center}
\resizebox{.6\hsize}{!}{\includegraphics{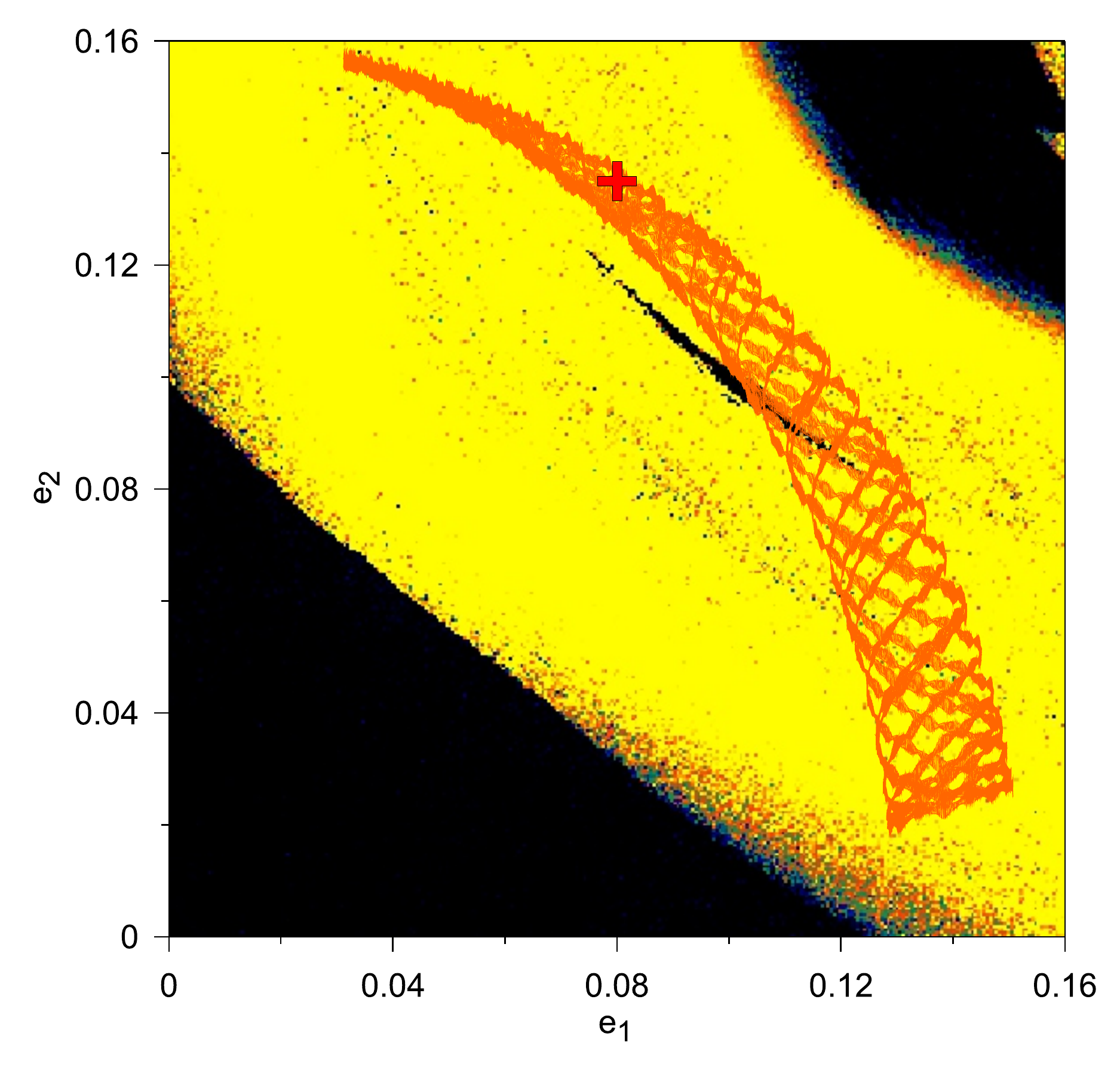}}\\
\resizebox{.4\hsize}{!}{\includegraphics[width=4.9cm,height=1.1cm]{bar.pdf}} 

\end{center}
\caption{DS map on $(e_1,e_2)$ plane guided by a spatial xz-symmetric periodic orbit. Although the observational eccentricities (red cross) of Kepler-108 lie in a chaotic domain, very close to them are the boundaries of a regular domain, which the orange evolution with $M_b=-110.226^{\circ}$ and $M_c=133.58^{\circ}$ surpasses.}
\label{kep108xz}
\end{figure}

Considering the respective mean anomalies for each configuration, in Fig. \ref{kep108fams} we overplot the evolutions based on the data of \citet{millsfab17} on the planar families in cyan (mean anomalies taken from the configuration (0,0)) and black (mean anomalies taken from the configuration $(0,\pi)$). Given the transit midpoints, we calculated the mean anomalies to $M_b=-110.226^{\circ}$ and $M_c=133.58^{\circ }$ , and the respective evolution is also presented (orange colour). We observe that all of them have similar evolutions and lie close to the v.c.o. (green and magenta dots) of the family in the configuration $(\theta_1,\theta_2)=(0,\pi)$. Let us recall that the v.c.o. are the orbits along a planar family of resonant periodic orbits where families of spatial periodic orbits bifurcate. In Fig.~\ref{kep108res}, the evolution for the system with mean anomalies $M_b=-110.226^{\circ}$ and $M_c=133.58^{\circ}$ (orange evolution in Fig. \ref{kep108fams}) is presented. It is shown that the resonant angles as well as the inclination-type resonant angles all rotate.

\begin{figure}
\centering
\resizebox{.6\hsize}{!}{\includegraphics{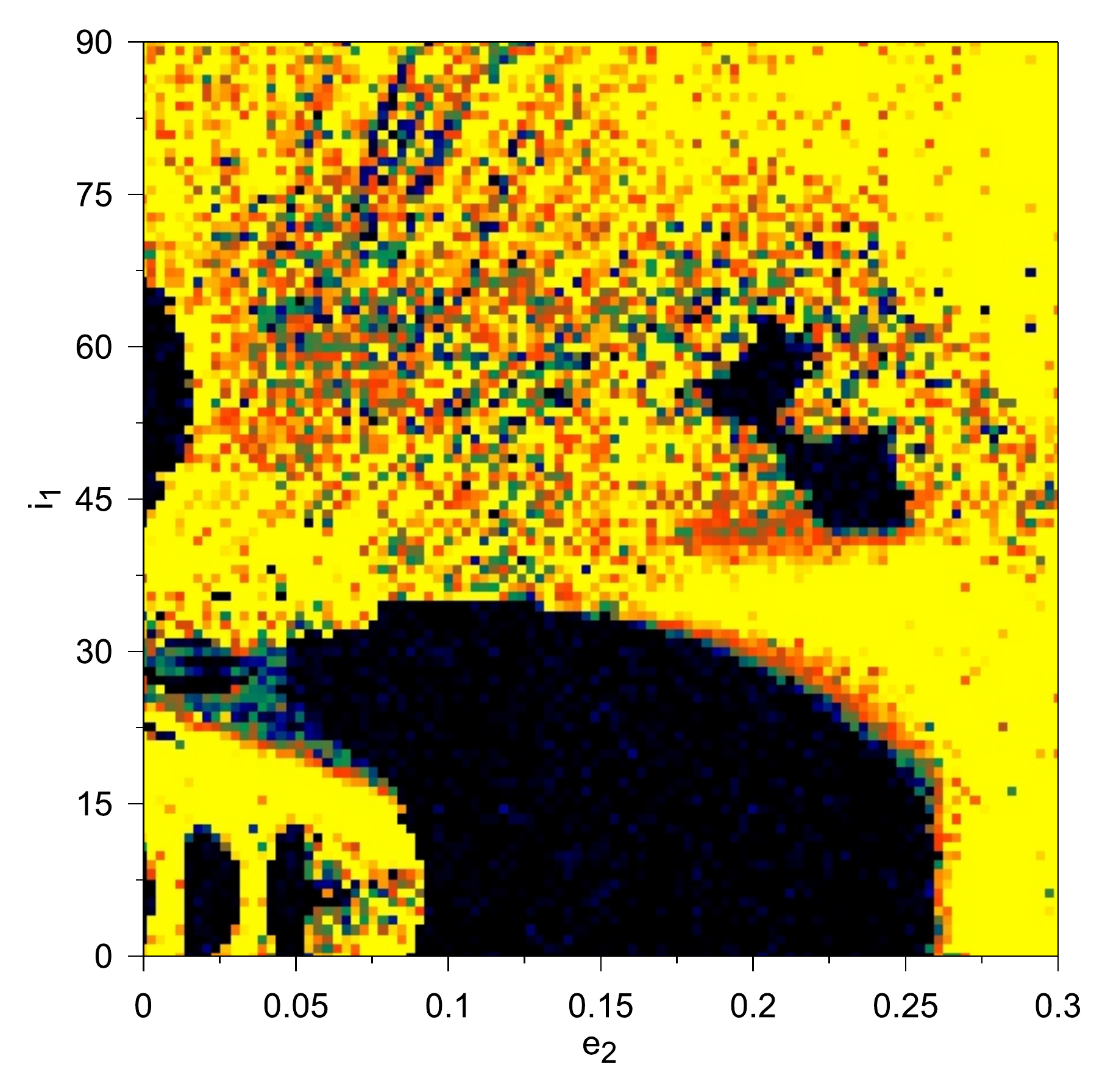}}

\resizebox{.6\hsize}{!}{\includegraphics{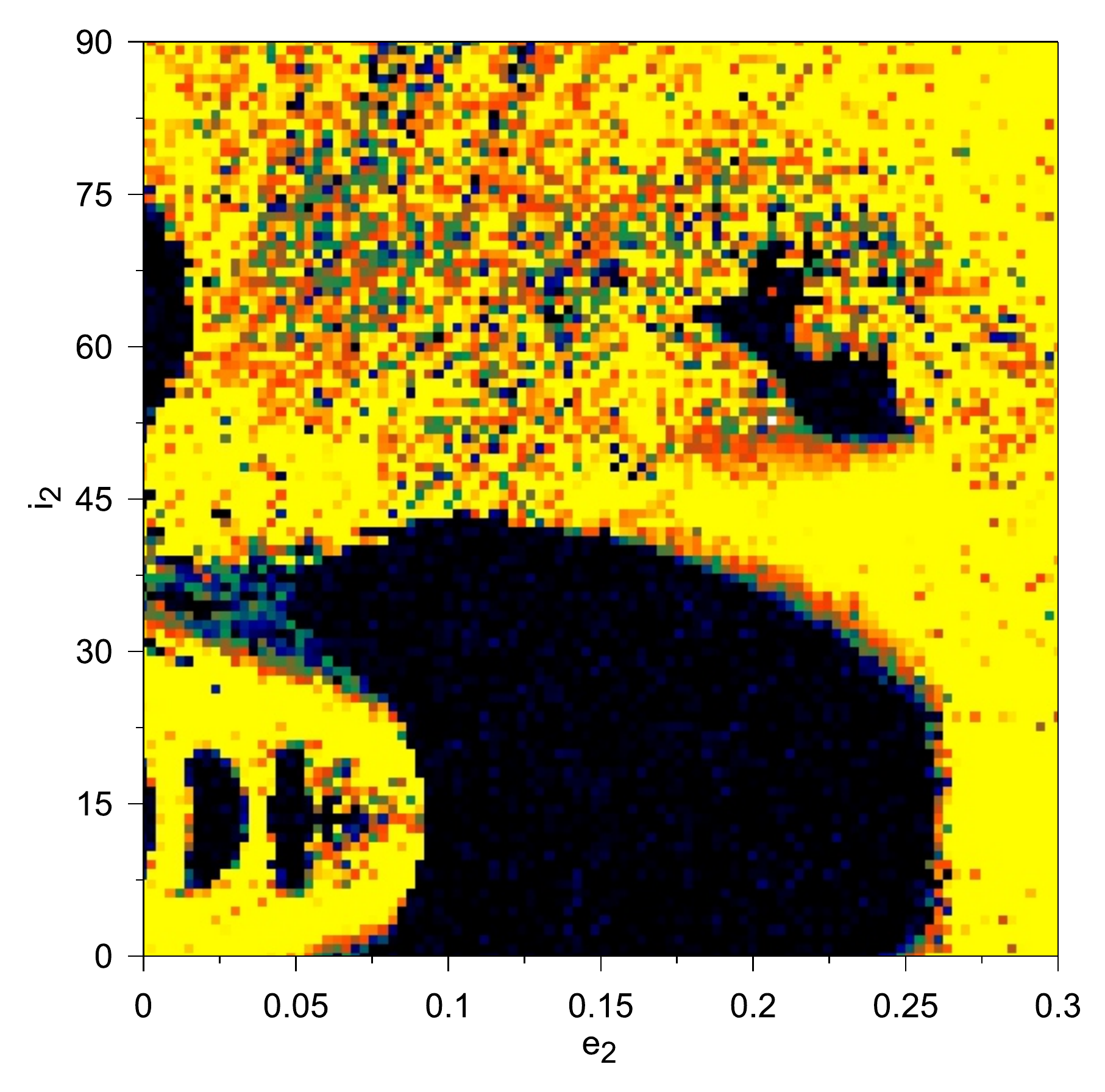}}

\resizebox{.4\hsize}{!}{\includegraphics[width=4.9cm,height=1.1cm]{bar.pdf}} 
\caption{DS map on $(e_2,i_1)$ (top) and $(e_2,i_2)$ (bottom) plane guided by a planar horizontally and vertically stable periodic orbit of the family in the configuration $(\theta_1,\theta_2)=(0,\pi)$ presented in Fig. \ref{kep108fams}.}
\label{kep108map}
\end{figure}

To identify a possible origin of the 3D system configuration favoured by \citet{millsfab17}, we computed the spatial families of periodic orbits (bifurcating from the green and magenta dots) and presented them in Fig. \ref{kep1083dfams} in the $(e_1,e_2, \Delta i)$ space. In the dynamical neighbourhood of the orange evolution with $M_b=-110.226^{\circ}$ and $M_c=133.58^{\circ}$ at $\Delta i=24^{\circ}$, there is a spatial family of stable (blue) xz-symmetric periodic orbits, which can guarantee the long-term stability of the mutually inclined planets.  To illustrate it, in Fig. \ref{kep108xz} we provide an example of the phase space through the DS map on the eccentricities plane in the vicinity of {the spatial periodic orbit that is crossed by the orange evolution in Fig. \ref{kep1083dfams}. The initial conditions are: $a_1/a_2=0.396383$, $i_1=18.38^{\circ}$, $i_2=5.55^{\circ}$, $\omega_1=\omega_2=90^{\circ}$, $M_1=0^{\circ}$, $M_2=180^{\circ}$, $\Omega_1=270^{\circ}$, and $\Omega_2=90^{\circ}$. The area around the planets (red cross) is chaotic, yet close to the boundaries of a regular domain. The orange evolution surpasses this narrow island of stability (created by the xz-symmetric spatial periodic orbit), where both the resonant angles and the inclination-type angles are librating, hence indicating an MMR and an inclination-type resonance that could guarantee the stability of mutually inclined planets.

Additionally, there is another possible neighbourhood that could host inclined planetary configurations. In order to reveal it, we constructed DS maps guided by the horizontally and vertically stable planar periodic orbits (solid blue lines) of Fig. \ref{kep108fams}. As an initial condition, we used the planar (namely, $i_1=i_2=0^{\circ}$) periodic orbit of the family in the configuration $(\theta_1,\theta_2)=(0,\pi)$, whose eccentricity value of the inner planet was the closest to $e_b$. This way, in Fig. \ref{kep108map} we kept the following fixed: $a_1/a_2=0.396401$, $e_1=0.080717$, $\varpi_1=M_1=0^{\circ}$, $\varpi_2=M_2=180^{\circ}$, and  $\Omega_1=\Omega_2=0$.  We varied the elements, ($i_1,e_2$) with $i_2$ being fixed to $0^{\circ}$,
and ($i_2,e_2$) with $i_1$ being fixed to $0^{\circ}$. We observe that the chosen periodic orbit could yield stability regions for $i_i<35^{\circ}$ ($i=1,2$), as shown by the dark regions of the DS maps in Fig. \ref{kep108map}.

Through all of these DS maps, we revealed the extent of two possible dynamical neighbourhoods that could justify the mutually inclined scenario favoured by \citet{millsfab17}. In the first case, the long-term stability at the mutual inclination of $\sim24^{\circ}$ could be maintained since the evolution surpasses the stable domain created by a spatial stable periodic orbit for $\Delta i=24^{\circ}$ , hence yielding a locking in an MMR and an inclination-type resonance inside it. In the second case, we showed that this mutual inclination is also within another island created by a planar periodic orbit ($\Delta i=i_1+i_2<35^{\circ}$) at $e_2=0.13515=e_c$ and $e_1=0.080717\approx e_b$, where we have a locking in an MMR.

\section{Conclusions}\label{con}
Many exoplanetary systems have been discovered to date that possess planets in MMR or resonant chains. The transit method coupled with TTV analysis provides an insight into the physical and orbital parameters of the detected systems, but suffers from observational limitations. 

Here, we show that additional validations or constraints on the orbital elements can be achieved through dynamical analyses based on periodic orbits computed for the GTBP. The intrinsic property of the periodic orbits, namely the linear horizontal and vertical stability, acts as a guide in the search of dynamical neighbourhoods that could host exoplanets. We showcased our dynamical approach for several planetary systems whose pair of planets is near an MMR. In particular, we applied the above-mentioned predictive power of periodic orbits to K2-21, K2-24, Kepler-9, and Kepler-108.

Our findings are as follows. Regarding the K2-21 system, there is a family of stable 5/3 resonant periodic orbits that could protect the planets, where the evolution is centred when the observational errors on the longitudes of pericentre are taken into account. Since this system is very sensitive with respect to the angles, in this study we provided the nominal ranges for the angles (mean anomalies included) that guarantee stability. 

We revealed that the two planets of the K2-24 system are protected by the 1/1 secondary resonance inside the 2/1 MMR. Thanks to our dynamical study, the eccentricities provided by \citet{petig18} were constrained even more ($e_2<0.05$).

Planets b and c of Kepler-9 evolve near the 2/1 MMR, but the family they reside close to consists of unstable periodic orbits. Our dynamical analysis highlights that this planetary architecture is viable, due to the apsidal difference oscillation that protects the phases of the planets and guarantees their survival.  

Kepler-108 is a system for which the non-coplanarity of the orbits is strongly envisaged \citep{millsfab17}. Being close to the 4/1 MMR, the planets seem to evolve in a stable region that stems from either a stable spatial periodic orbit or a vertically stable planar periodic orbit. The value for the proposed mutual inclination of $24^{\circ}$ lies well within these regular domains.

Since the long-term evolution and stability of planetary systems are crucial for their habitability, our work exemplifies the need for dynamical analyses based on periodic orbits for the optimum deduction of the orbital elements of newly discovered exoplanets. In the future, such a study could be even more efficient during the fitting process of the observational data.

\vspace{0.2cm}
{\bf Acknowledgements}
We thank the anonymous reviewer whose remarks helped us improve our manuscript. The research of KIA is co-financed by Greece and the European Union (European Social Fund- ESF) through the Operational Programme \guillemotleft Human Resources Development, Education and Lifelong Learning \guillemotright in the context of the project ``Reinforcement of Postdoctoral Researchers - $2^{\rm{nd}}$ Cycle'' (MIS-5033021), implemented by the State Scholarships Foundation (IKY). This work was supported by the Fonds de la Recherche Scientifique - FNRS under Grant No. F.4523.20 (DYNAMITE MIS-project). Computational resources have been provided by the Consortium des \'Equipements de Calcul Intensif, supported  by the FNRS-FRFC, the Walloon Region, and the University of Namur (Conventions No. 2.5020.11, GEQ U.G006.15, 1610468 et RW/GEQ2016).

\bibliographystyle{aa} 

\bibliography{ttvposab} 

\end{document}